\documentstyle[preprint,aps,epsfig]{revtex}
\tightenlines
\begin{document}
\draft
\preprint{\vbox{
\hbox{IFT-P.047/2000}
\hbox{May 2000}
}}
\title{MSW effect with flavor changing
interactions and the atmospheric neutrino problem
} 
\author{ J. C. Montero~\footnote{e-mail address:montero@ift.unesp.br} and 
V. Pleitez~\footnote{e-mail address: vicente@ift.unesp.br}}  
\address{
Instituto de F\'\i sica Te\'orica\\
Universidade Estadual Paulista\\
Rua Pamplona 145\\ 
01405-900-- S\~ao Paulo, SP\\Brazil} 
\date{\today}
\maketitle
\begin{abstract}

We consider flavor changing effective neutrino interactions
in the context of massive neutrinos in the issue of atmospheric neutrinos.
Assuming as usual that this is an indication of the oscillation of muon 
neutrinos into tau neutrinos we show that there is a set of 
parameters which is consistent with the MSW resonance condition for the
typical Earth density and atmospheric neutrino energies. In particular
we show that even if the vacuum mixing angle vanishes   
it is still possible to have a resonance which is compatible with the 
atmospheric neutrino data. 
We also briefly consider the case of the solar neutrino problem.

\end{abstract}

It is a well accepted fact that the solar~\cite{solar} and atmospheric
neutrino data~\cite{sk1} strongly suggest that neutrinos are massive 
particles~\cite{vb}.
Other solutions like pure flavor changing neutrino 
interactions~\cite{fcnit} or neutrino magnetic 
moments~\cite{mg1,mg2} are still possible 
for the solar neutrino problem, at least with the present data. 
Although the more favoured solution is the vacuum oscillation one, it is 
still possible that two or more of these mechanisms are in 
action simultaneously. For instance, the MSW~\cite{lw1,ms} effect together 
with effective flavor changing neutrino interactions was proposed in 
Ref.~ \cite{er}.
This is an interesting possibility since almost all extensions of the 
electroweak standard model implies new interactions for the neutrinos. 

On the other hand, several years ago Bethe~\cite{bethe} re-derived the 
resonance effect on neutrinos propagating through a medium~\cite{lw1,ms}
based on a different view: his approach consists on considering the 
addition of a term $2EV$ upon the neutrino mass square due to the 
$W$-interactions of massive neutrinos propagating in a medium;  
$E$ is the neutrino energy and $V$ 
is the potential energy due to the matter effect.
Here we will consider a similar effect  
when new interactions with matter (electrons or quarks) 
do exist and also if the neutrinos have arbitrary masses and for the sake 
of simplicity we will treat only two generations. 

In this situation, the neutrino symmetry eigenstates $\nu_\alpha$ 
and $\nu_\beta$ are related to the neutrino mass eigenstates $\nu_1$ 
and $\nu_2$ as follows
\begin{equation}
\vert \nu_\alpha\rangle=
\vert\nu_1\rangle\cos\theta+\vert\nu_2\rangle\sin\theta,
\quad
\vert \nu_\beta\rangle=
-\vert\nu_1\rangle\sin\theta+\vert\nu_2\rangle\cos\theta
\label{nutau}
\end{equation}
and $\theta<45^o$; the masses of the states $\vert\nu_{1,2}\rangle$ are 
$m_{1,2}$ respectively and as we said before, we are assuming that they are 
arbitrary parameters.
The square of the mass matrix in the $\vert\nu_{\alpha,\beta}\rangle$ 
flavor basis is given by

\begin{equation}
{\cal M}^2=\frac{1}{2}(m^2_1+m^2_2)\left(
\begin{array}{cc}
1& 0\\
0&1
\end{array}
\right)+\frac{1}{2}(m^2_2-m^2_1)
\left(
\begin{array}{cc}
-\cos2\theta & \sin2\theta\\
\sin2\theta&\cos2\theta
\end{array}
\right).
\label{mm}
\end{equation}

Assuming effective interactions of the form  
$(G_{\alpha f}G_{f\beta}/M^2_X)(\bar{\nu}_\alpha\nu_\beta)(\bar{f}f),
\;\alpha,\beta=e,\mu,\tau$; and $M_X$ is the mass of the interchanged particle 
and $f$ denotes a charged lepton or quark 
(we have omitted Lorentz
indices) implies a potential energy that is given by  
(here we will consider only interactions with the electrons i.e., $f=e$): 
\begin{equation}
V^{NI}_{\alpha\beta}=
\frac{x}{M^2_W}
\,G_{\alpha e}G_{e \beta}N_e,\quad \alpha,\beta=e,\mu,\tau;
\label{inteff}
\end{equation}
with $x=M^2_W/M^2_X$; $N_e$ is the number of electrons per unit volume. 
Constraints on $x$ and $G_{\alpha\beta}$ coming from several decays are 
model dependent so they will not be considered here at all. 
However, see Ref.~\cite{fcnit}.

In fact, our considerations are model independent but neutrinos must be 
massive particles with renormalizable (arbitrary) masses.
This is a necessary condition in order   
to have neutrino masses that do not depend on the parameters of
the flavor changing interactions $G_{\alpha e}$. 
For instance, in the context of the $R$-parity broken MSSM~\cite{rpv} it 
means that one has to add right-handed neutrinos and its respective 
sneutrinos $\tilde{\nu}_R$. In the so called 3-3-1 model~\cite{331} 
neutrinos can have arbitrary Majorana masses if a neutral component of 
the sextet gains a non vanishing vacuum expectation value~\cite{ma1}, or 
we can also add three right-handed neutrinos (and Dirac neutrinos), or 
both possibilities. 

The momentum of a neutrino is related to its energy $E$ by
\begin{equation}
k^2+m^2=(E-V)^2\approx E^2-2VE,
\label{mo}
\end{equation}
and considering the atmospheric neutrino issue i.e., the muon and tau neutrino 
we see that $V_{\alpha\beta}$ is equivalent to an addition to $m^2$ given 
by
\begin{equation}
m^2_{\alpha\beta}=2V_{\alpha\beta}E=2\sqrt{2}\,
\left(\frac{x}{\sqrt2}\,G_{\alpha e}G_{e \beta }
 +G_FM^2_W(1-4s^2_W)\,\delta_{\alpha\beta}\right)
\frac{Y_e}{M^2_Wm_n}\,\rho E
\equiv A_{\alpha\beta}
\label{m2}
\end{equation}
where $\alpha,\beta=\mu,\tau$; $m_n$ is  the mass of the nucleon and 
$Y_e$ the number of electrons per
nucleon in the matter, usually $Y_e=1/2$ and we have introduced the $Z^0$ 
contribution as well. Taking the usual values for the parameters~\cite{pdg} 
we have:

\begin{equation}
A_{\alpha\beta}=1.0\times10^{-3}\,\left(\frac{x}{\sqrt2}\,
G_{\alpha e}G_{e\beta }+0.006\,\delta_{\alpha\beta}
\right)\,\rho E,
\label{as}
\end{equation}
where $\rho$ is in grams per cubic centimeter, $E$ is in GeV, and 
$A_{\alpha\beta}$ is in $\mbox{eV}^2$.

Thus, in a dense medium the original vacuum mass square matrix is modified
by the $A_{\alpha\beta}$ term and it reads:
\begin{equation}
\tilde{{\cal M}}^2=\frac{1}{2}\left(m^2_1+m^2_2+A_{\mu\mu}+A_{\tau\tau}\right)
\left(
\begin{array}{cc}
1& 0\\
0&1
\end{array}
\right)+\frac{1}{2}
\left(
\begin{array}{cc}
A_{\mu\mu}-A_{\tau\tau}-\Delta\cos2\theta & 2A_{\mu\tau}+\Delta\sin2\theta\\
2A_{\mu\tau}+\Delta\sin2\theta& -A_{\mu\mu}+A_{\tau\tau}+\Delta\cos2\theta
\end{array}
\right).
\label{mmi}
\end{equation}
where $\Delta=m^2_2-m^2_1$ and we will assume $m^2_2>m^2_1$. 
The eigenvalues are

\begin{equation}
\tilde{m}^2_{1,2}=
\frac{1}{2}\left(m^2_1+m^2_2+A_{\mu\mu}+A_{\tau\tau}\right)\pm
\frac{1}{2}\left[(-A_{\mu\mu}+A_{\tau\tau}+
\Delta\cos2\theta)^2+\vert 2A_{\mu\tau}+ \Delta\sin2\theta\vert^2\right]^{1/2},
\label{eigenvalues}
\end{equation}
and we see that the resonance condition is given by:
\begin{equation}
A_{\mu\mu}-A_{\tau\tau}=
\Delta\cos2\theta.
\label{rc}
\end{equation}

\begin{figure}[ht]
\begin{center}
\centering\leavevmode
\epsfxsize=\hsize
\epsfbox{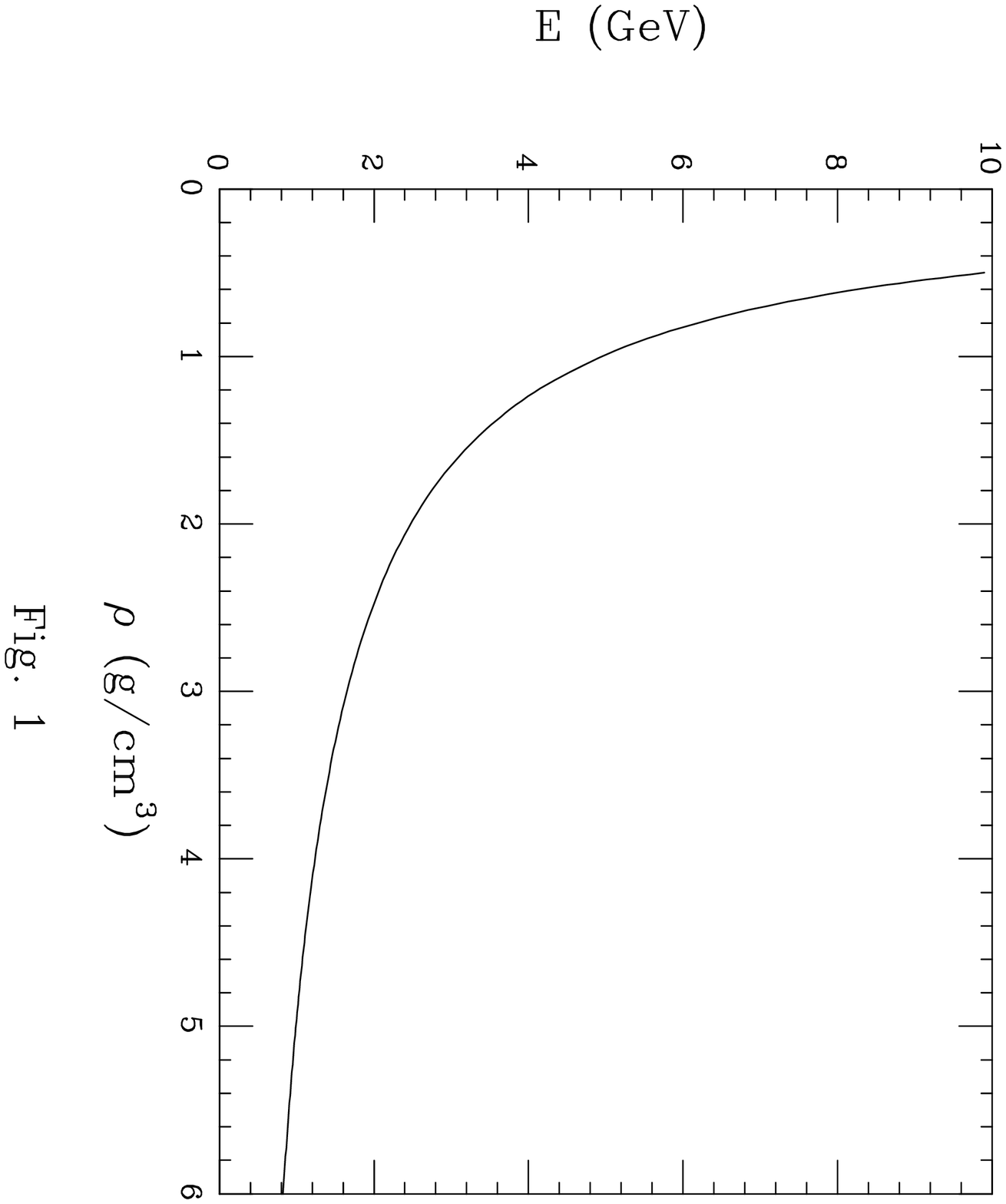}
\vglue -0.009cm
\end{center}
\caption{The energy of the neutrinos satisfying the resonance condition
as a function of the Earth density for $G_{\mu e}=0.9,\,G_{\tau e }=0.25,\,
\Delta=2.2\times10^{-3}\mbox{eV}^2$ and $\cos2\theta=1$.}
\label{fig1}
\end{figure}

\begin{figure}[ht]
\begin{center}
\centering\leavevmode
\epsfxsize=400pt
\epsfbox{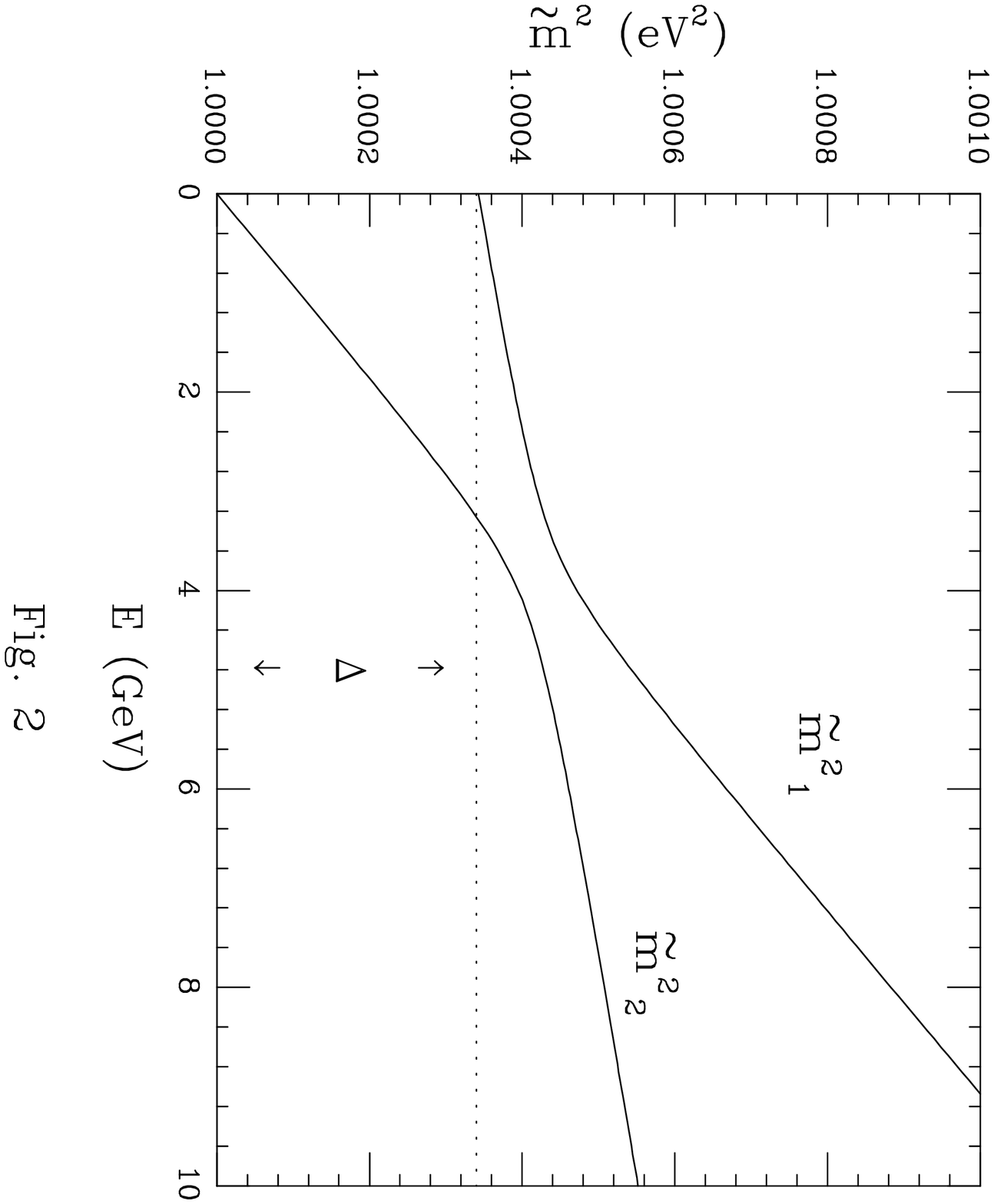}
\vglue -0.009cm
\end{center}
\vglue 2cm
\caption{ The quantities $\tilde{m}^2_{1,2}$ as functions of the neutrino 
energy for $\rho=2\;\mbox{g}/\mbox{cm}^3$, $m^2_1=1\,\mbox{ eV}^2, 
\;m^2_2=m^2_1+\Delta$ with $\Delta=3.4\times10^{-3}\,\mbox{eV}^2$ and 
$\cos2\theta=1$.} 
\label{fig2}
\end{figure}

In Fig.~\ref{fig1} we show the 
neutrino energy $E$ as a function of the
Earth density at the resonance given by Eq.~(\ref{rc}) for a given set of 
parameters, in particular with vanishing vacuum mixing angle.
In Fig.~\ref{fig2} we show the two eigenvalues
$\tilde{m}^2_{1,2}$ as a function of $E$ and not as a function of $\rho$
and we see that a conversion is possible for the typical Earth density and
neutrino energies also with $\cos2\theta=1$. The occurrence of this resonance
with $\theta=0$ does not occur in the context of the standard model.
Both figures only illustrate that typical values for the Earth density
and atmospheric neutrino energies are compatible with the resonance
picture discussed above.  

In fact, we can also consider the survival probability in a medium which is 
given by
\begin{equation}
P(\nu_\mu\to\nu_\mu)=1-\sin^22\tilde\theta\sin^2\left(1.27\,
\tilde\Delta\,\frac{L}{E}\right),
\label{nunu}
\end{equation}
where 
\begin{equation}
\sin^22\tilde\theta=\frac{(2A_{\mu\tau}+\Delta\sin2\theta)^2}
{(\Delta\cos2\theta-A_{\mu\mu}+A_{\tau\tau})^2+
(2A_{\mu\tau}+\Delta\sin2\theta)^2},\quad \mbox{and}
\quad \tilde\Delta=\tilde{m}^2_2-\tilde{m}^2_1,
\label{def}
\end{equation}
with $\tilde\Delta$ in $\mbox{eV}^2$, $L$ in km and $E\approx \tilde{E}=
\vert\vec{p}\vert$ in 
GeV; and we have already used $\tilde{E}_2-\tilde{E}_1\approx 
\tilde\Delta/2\vert\vec{p}\vert$.
 
Notice that we can explain the atmospheric neutrino data 
since at the resonance $\sin^22\tilde\theta_r=1$ and if 
$\tilde\Delta_r=2.2\times10^{-3}\,\mbox{eV}^2$ (the subscript $r$ denotes 
the value at the resonance), which corresponds to the best 
fit of atmospheric neutrinos~\cite{sk1,sk2} leaving the vacuum parameters
$\theta$ and $\Delta$ arbitrary. We have

\begin{equation}
\tilde\Delta^2=(\tilde{m}^2_2-\tilde{m}^2_1)^2=(-A_{\mu\mu}+A_{\tau\tau}+
\Delta\cos2\theta)^2+\vert 2A_{\mu\tau}+ \Delta\sin2\theta\vert^2,
\label{yupi}
\end{equation}
and we see that, even at the resonance,  if $\theta=0$, we can still
have that $\tilde\Delta_r=2\vert A_{\mu\tau,r}\vert=2.2\times10^{-3}
\mbox{eV}^2$, which is accomplished, for instance, for neutrinos with
$E\approx2$ GeV by taken $G_{\mu e}\approx G_{\tau e}\approx1,\,\rho=
2\,\mbox{g}/\mbox{cm}^3$. 
This also occurs if $\Delta=0$ (neutrinos mass degenerated or massless 
neutrinos) which also implies $A_{\mu\mu}=
A_{\tau\tau}$ at the resonance. 
This case corresponds to the pure flavor changing induced 
neutrino oscillations as in Ref.~\cite{rpatm} but this situation
may have problems with the up-going upward muon~\cite{sk2} as has been
pointed out in Ref.~\cite{ln}. 

Likewise, we can consider the solar neutrino problem. In this case we
have to change labels,
$\mu\to e$, $\tau\to \mu$, in the expressions above and also to add
the contribution of the $W$ boson. Then

\begin{equation}
m^2_{\alpha\beta}=2V_{\alpha\beta}E=2\sqrt{2}\,
\left(\frac{x}{\sqrt2}\,G_{\alpha e}G_{e\beta}
 +G_FM^2_W\,[\delta_{\alpha e}\delta_{\beta e}+(1-4s^2_W)
\delta_{\alpha\beta}]\right)
\frac{Y_e}{M^2_Wm_n}\,\rho E
\equiv A_{\alpha\beta},
\label{m3}
\end{equation}
for $\alpha,\beta=e,\mu$. Putting values we have

\begin{equation}
A_{\alpha\beta}=1.0\times10^{-3}\,\left(\frac{x}{\sqrt2}\,
G_{\alpha e}G_{e\beta}+ 0.075\,\delta_{\alpha e}\delta_{\beta e}+
0.006\,\delta_{\alpha\beta}\right)\,\rho E,
\label{as2}
\end{equation}
with the same units as in Eq.~(\ref{as}).

\begin{figure}[ht]
\begin{center}
\centering\leavevmode
\epsfxsize=400pt
\epsfbox{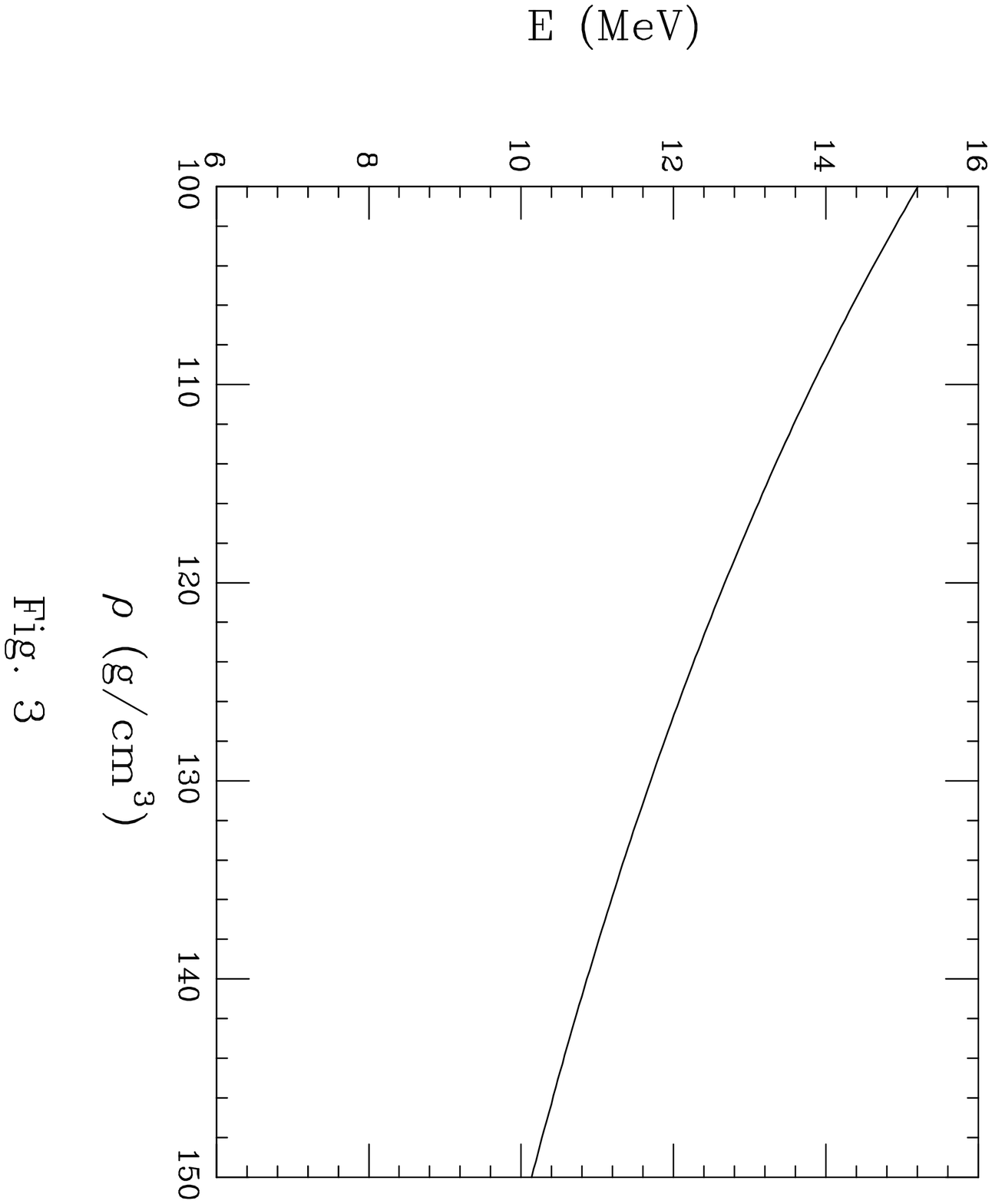}
\vglue -0.009cm
\end{center}
\vglue 2cm
\caption{ The same as in Fig.~1 with the Sun density and $G_{ee}=1.0,\,
G_{\mu e}=0.9,\,\Delta=2.0\times10^{-4}\mbox{eV}^2$ and $\cos2\theta=1$. }
\label{fig3}
\end{figure}

The eigenvalues are

\begin{equation}
\tilde{m}^2_{1,2}=
\frac{1}{2}\left(m^2_1+m^2_2+A_{ee}+A_{\mu\mu}\right)\pm
\frac{1}{2}\left[(-A_{ee}+A_{\mu\mu}+
\Delta\cos2\theta)^2+\vert 2A_{e\mu}+ \Delta\sin2\theta\vert^2\right]^{1/2}.
\label{eigenvalues2}
\end{equation}
$\Delta$ has the same definition as before, however the masses $m_{1,2}$ 
may be different from that of the muon and tau neutrino case.
In this case we have a similar resonance condition than that in 
Eq.~(\ref{rc}). In Fig.~\ref{fig3} we show neutrino energies as
a function of typical Sun density at the resonance for a given set of
parameters and using $\cos2\theta=1$ again.  
The occurrence of that resonance at $\theta=0$ does not happens 
if we consider the electron neutrino and the muon neutrino as in the Bethe's 
paper~\cite{bethe}. The latter case is got here if $A_{ee}=A$, 
$A_{\mu\mu}=A_{e\mu}=0$ and $\theta\not=0$.  

Since in the present context the atmospheric neutrino data is almost 
independent of the value of vacuum mass square difference,
it means that we could be able to explain 
solar~\cite{solar}, atmospheric~\cite{sk1,sk2} and LSND~\cite{lsnd} neutrino
data with three active neutrinos and without introducing a sterile neutrino. 
This will be the case if, in a truly three generation case, the 
$\Delta_{12}$ which dominates the 
$\nu_\mu\to\nu_\tau$ oscillation also dominates the $\nu_e\to\nu_\tau$
transition; while $\Delta_{13}$ is the one used for fitting the LSND results.

We have worked in the two-neutrino scheme because most of the data
are presented in this situation. However a general three-neutrino scheme
must be worked elsewhere; also more realistic fittings by considering the
matter density profile of the Earth and Sun eventually have to be 
done~\cite{mmg}.

\acknowledgments 
This work was supported by Funda\c{c}\~ao de Amparo \`a Pesquisa
do Estado de S\~ao Paulo (FAPESP), Conselho Nacional de 
Ci\^encia e Tecnologia (CNPq) and by Programa de Apoio a
N\'ucleos de Excel\^encia (PRONEX).

\vglue 0.01cm

\vglue 0.01cm

\vglue 0.01cm
\end{document}